\documentclass[12pt,a4paper]{article}

\usepackage{ifthen} 
\newboolean{pdflatex}
\setboolean{pdflatex}{true} 

\newboolean{articletitles}
\setboolean{articletitles}{true} 

\newboolean{uprightparticles}
\setboolean{uprightparticles}{false} 

\def\paperauthors{LHCb collaboration} 
\def\paperasciititle{Study of exclusive photoproduction of charmonium in ultra-peripheral lead-lead collisions} 
\def\papertitle{Study of exclusive photoproduction of charmonium in ultra-peripheral lead-lead collisions} 
\def\paperkeywords{{High Energy Physics}, {LHCb}} 
\def\papercopyright{\the\year\ CERN for the benefit of the LHCb collaboration} 
\def\paperlicence{CC-BY-4.0 licence}
\def\paperlicenceurl{https://creativecommons.org/licenses/by/4.0/}

\usepackage[top=1in, bottom=1.25in, left=1in, right=1in]{geometry}

\usepackage{microtype}
\usepackage{lineno}  
\usepackage{xspace} 
\usepackage{caption} 

\usepackage{graphicx}  
\usepackage{color}
\usepackage{colortbl}
\graphicspath{{./figs/}} 

\usepackage{amsmath} 
\usepackage{amssymb}
\usepackage{amsfonts}
\usepackage{upgreek} 

\newcommand*\patchAmsMathEnvironmentForLineno[1]{%
\expandafter\let\csname old#1\expandafter\endcsname\csname #1\endcsname
\expandafter\let\csname oldend#1\expandafter\endcsname\csname
end#1\endcsname
 \renewenvironment{#1}%
   {\linenomath\csname old#1\endcsname}%
   {\csname oldend#1\endcsname\endlinenomath}%
}
\newcommand*\patchBothAmsMathEnvironmentsForLineno[1]{%
  \patchAmsMathEnvironmentForLineno{#1}%
  \patchAmsMathEnvironmentForLineno{#1*}%
}
\AtBeginDocument{%
\patchBothAmsMathEnvironmentsForLineno{equation}%
\patchBothAmsMathEnvironmentsForLineno{align}%
\patchBothAmsMathEnvironmentsForLineno{flalign}%
\patchBothAmsMathEnvironmentsForLineno{alignat}%
\patchBothAmsMathEnvironmentsForLineno{gather}%
\patchBothAmsMathEnvironmentsForLineno{multline}%
\patchBothAmsMathEnvironmentsForLineno{eqnarray}%
}

\usepackage{hyperxmp}

\usepackage[pdftex,
            pdfauthor={\paperauthors},
            pdftitle={\paperasciititle},
            pdfkeywords={\paperkeywords},
            pdfcopyright={Copyright (C) \papercopyright},
            pdflicenseurl={\paperlicenceurl}]{hyperref}

\usepackage[colorinlistoftodos,textsize=scriptsize]{todonotes}

\usepackage[bottom,flushmargin,hang,multiple]{footmisc}

\usepackage[all]{hypcap} 

\usepackage{xspace} 
\usepackage{upgreek}

\def\tot{\mathrm{tot}}
\def\coh{\mathrm{coh}}
\def\logpt2 {\ensuremath{\ln(\pt^{*2})}\xspace} 
\def\ptsq {\ensuremath{\pt^{*2}}\xspace} 

\def\gevcsquare    {\ensuremath{{\,(\mathrm{Ge\kern -0.1em V\!/}c)^2}}\xspace}
\def\mygevc        {\ensuremath{{\mathrm{Ge\kern -0.1em V\!/}c}}\xspace}
\def\unitmevc   {\ensuremath{\mathrm{[Me\kern -0.1em V\!/}c]}\xspace}
\def\unitgevc   {\ensuremath{\mathrm{[Ge\kern -0.1em V\!/}c]}\xspace}
\def\unitmbarngevc{\ensuremath{[\mathrm{mb/(Ge\kern -0.1em V\!/}c)]}\xspace}
\def\unitmbarn{\ensuremath{[\mathrm{mb}]}\xspace}

\def\lhcb   {\mbox{LHCb}\xspace}

\def\cms    {\mbox{CMS}\xspace}
\def\alice  {\mbox{ALICE}\xspace}

\def\lhc    {\mbox{LHC}\xspace}

\def\herschel {\mbox{\textsc{HeRSCheL}}\xspace}
\def\MagUp {\mbox{\em Mag\kern -0.05em Up}\xspace}

\ifthenelse{\boolean{uprightparticles}}%
{

 \def\Pmu         {\ensuremath{\upmu}\xspace}

 \def\Ppsi        {\ensuremath{\uppsi}\xspace}

 \def\PDelta      {\ensuremath{\Delta}\xspace}                 
 \def\PXi         {\ensuremath{\Xi}\xspace}                 
 \def\PLambda     {\ensuremath{\Lambda}\xspace}                 
 \def\PSigma      {\ensuremath{\Sigma}\xspace}                 
 \def\POmega      {\ensuremath{\Omega}\xspace}                 
 \def\PUpsilon    {\ensuremath{\Upsilon}\xspace}

 \def\PB      {\ensuremath{\mathrm{B}}\xspace}                 
                  
 \def\PD      {\ensuremath{\mathrm{D}}\xspace}

 \def\PJ      {\ensuremath{\mathrm{J}}\xspace}                 
 \def\PK      {\ensuremath{\mathrm{K}}\xspace}

 \def\Pb      {\ensuremath{\mathrm{b}}\xspace}                 
 \def\Pc      {\ensuremath{\mathrm{c}}\xspace}                 
                  
 \def\Pe      {\ensuremath{\mathrm{e}}\xspace}

 \def\Pi      {\ensuremath{\mathrm{i}}\xspace}

 \def\Ps      {\ensuremath{\mathrm{s}}\xspace}

 \def\thebaroffset{0.0em}
}
{

 \def\Pmu         {\ensuremath{\mu}\xspace}

 \def\Ppsi        {\ensuremath{\psi}\xspace}                 
                  
 \mathchardef\PDelta="7101
 \mathchardef\PXi="7104
 \mathchardef\PLambda="7103
 \mathchardef\PSigma="7106
 \mathchardef\POmega="710A
 \mathchardef\PUpsilon="7107
                  
 \def\PB      {\ensuremath{B}\xspace}                 
                  
 \def\PD      {\ensuremath{D}\xspace}

 \def\PJ      {\ensuremath{J}\xspace}                 
 \def\PK      {\ensuremath{K}\xspace}

 \def\Pb      {\ensuremath{b}\xspace}                 
 \def\Pc      {\ensuremath{c}\xspace}                 
                  
 \def\Pe      {\ensuremath{e}\xspace}

 \def\Pi      {\ensuremath{i}\xspace}

 \def\Ps      {\ensuremath{s}\xspace}

 \def\thebaroffset{0.18em}
}
\newcommand{\offsetoverline}[2][\thebaroffset]{\kern #1\overline{\kern -#1 #2}}%

\makeatletter
\ifcase \@ptsize \relax
  \newcommand{\miniscule}{\@setfontsize\miniscule{4}{5}}
\or
  \newcommand{\miniscule}{\@setfontsize\miniscule{5}{6}}
\or
  \newcommand{\miniscule}{\@setfontsize\miniscule{5}{6}}
\fi
\makeatother

\DeclareRobustCommand{\optbar}[1]{\shortstack{{\miniscule (\rule[.5ex]{1.25em}{.18mm})}
  \\ [-.7ex] $#1$}}

\def\epem       {{\ensuremath{\Pe^+\Pe^-}}\xspace}

\def\mumu       {{\ensuremath{\Pmu^+\Pmu^-}}\xspace}

\def\squark    {{\ensuremath{\Ps}}\xspace}

\def\cquark    {{\ensuremath{\Pc}}\xspace}

\def\bquark    {{\ensuremath{\Pb}}\xspace}

\def\KorKbar {\kern \thebaroffset\optbar{\kern -\thebaroffset \PK}{}\xspace}

\def\DorDbar {\kern \thebaroffset\optbar{\kern -\thebaroffset \PD}\xspace}

\def\B       {{\ensuremath{\PB}}\xspace}

\def\BorBbar {\kern \thebaroffset\optbar{\kern -\thebaroffset \PB}\xspace}

\def\Bd      {{\ensuremath{\B^0}}\xspace}

\def\BdorBdbar {\kern \thebaroffset\optbar{\kern -\thebaroffset \Bd}\xspace}

\def\Bs      {{\ensuremath{\B^0_\squark}}\xspace}

\def\BsorBsbar {\kern \thebaroffset\optbar{\kern -\thebaroffset \Bs}\xspace}

\def\jpsi     {{\ensuremath{{\PJ\mskip -3mu/\mskip -2mu\Ppsi\mskip 2mu}}}\xspace}
\def\psitwos  {{\ensuremath{\Ppsi{(2S)}}}\xspace}

\def\Y#1S{\ensuremath{\PUpsilon{(#1S)}}\xspace}

\def\LorLbar     {\kern \thebaroffset\optbar{\kern -\thebaroffset \PLambda}\xspace}

\def\BF         {{\ensuremath{\mathcal{B}}}\xspace}
\def\BR         {\BF}

\newcommand{\decay}[2]{\ensuremath{#1\!\to #2}\xspace} 

\def\to                 {\ensuremath{\rightarrow}\xspace}

\newcommand{\etot}{{\ensuremath{\varepsilon_{\mathrm{ tot}}}}\xspace}

\def\AT#1     {\ensuremath{A_{\mathrm{T}}^{#1}}\xspace}           

\def\C#1      {\ensuremath{\mathcal{C}_{#1}}\xspace}                       
\def\Cp#1     {\ensuremath{\mathcal{C}_{#1}^{'}}\xspace}                    
\def\Ceff#1   {\ensuremath{\mathcal{C}_{#1}^{\mathrm{(eff)}}}\xspace}        
\def\Cpeff#1  {\ensuremath{\mathcal{C}_{#1}^{'\mathrm{(eff)}}}\xspace}       
\def\Ope#1    {\ensuremath{\mathcal{O}_{#1}}\xspace}                       
\def\Opep#1   {\ensuremath{\mathcal{O}_{#1}^{'}}\xspace}                    

\newcommand{\nospaceunit}[1]{\ensuremath{\text{#1}}}       
\newcommand{\aunit}[1]{\ensuremath{\text{\,#1}}}       

\newcommand{\tev}{\aunit{Te\kern -0.1em V}\xspace}
\newcommand{\gev}{\aunit{Ge\kern -0.1em V}\xspace}
\newcommand{\mev}{\aunit{Me\kern -0.1em V}\xspace}
\newcommand{\kev}{\aunit{ke\kern -0.1em V}\xspace}
\newcommand{\ev}{\aunit{e\kern -0.1em V}\xspace}
\newcommand{\mevc}{\ensuremath{\aunit{Me\kern -0.1em V\!/}c}\xspace}
\newcommand{\gevc}{\ensuremath{\aunit{Ge\kern -0.1em V\!/}c}\xspace}
\newcommand{\mevcc}{\ensuremath{\aunit{Me\kern -0.1em V\!/}c^2}\xspace}
\newcommand{\gevcc}{\ensuremath{\aunit{Ge\kern -0.1em V\!/}c^2}\xspace}

\def\m    {\aunit{m}\xspace}

\def\mum  {\ensuremath{\,\upmu\nospaceunit{m}}\xspace}

\def\mbarn{\aunit{mb}\xspace}
\def\mub{\ensuremath{\,\upmu\nospaceunit{b}}\xspace}

\def\deriv {\ensuremath{\mathrm{d}}}

\def\gsim{{~\raise.15em\hbox{$>$}\kern-.85em
          \lower.35em\hbox{$\sim$}~}\xspace}
\def\lsim{{~\raise.15em\hbox{$<$}\kern-.85em
          \lower.35em\hbox{$\sim$}~}\xspace}

\def\sqsnn {\ensuremath{\protect\sqrt{s_{\scriptscriptstyle\text{NN}}}}\xspace}
\def\pt         {\ensuremath{p_{\mathrm{T}}}\xspace}
\def\ptsq       {\ensuremath{p_{\mathrm{T}}^2}\xspace}
\def\ptot       {\ensuremath{p}\xspace}

\newcommand{\lum} {\ensuremath{\mathcal{L}}\xspace}

\def\evtgen     {\mbox{\textsc{EvtGen}}\xspace}

\def\geant      {\mbox{\textsc{Geant4}}\xspace}

\def\photos     {\mbox{\textsc{Photos}}\xspace}

\def\tell1  {TELL1\xspace}
\def\ukl1   {UKL1\xspace}

\newcommand{\lhcborcid}[1]{\href{https://orcid.org/#1}{\hspace*{0.1em}\raisebox{-0.45ex}{\includegraphics[width=1em]{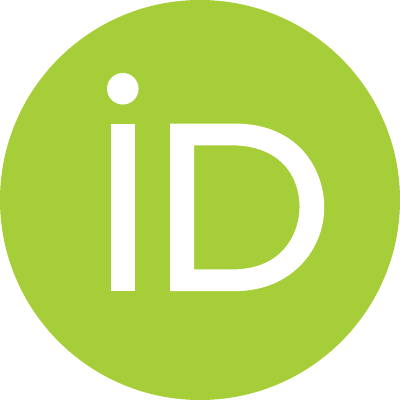}}}}

\usepackage{cite} 
\usepackage{mciteplus}

\usepackage{longtable} 
\usepackage{multirow}
\usepackage[utf8]{inputenc}

\usepackage{xcolor}
\usepackage[normalem]{ulem}

\begin{document}

\renewcommand{\thefootnote}{\fnsymbol{footnote}}
\setcounter{footnote}{1}


\begin{titlepage}
\pagenumbering{roman}

\vspace*{-1.5cm}
\centerline{\large EUROPEAN ORGANIZATION FOR NUCLEAR RESEARCH (CERN)}
\vspace*{1.5cm}
\noindent
\begin{tabular*}{\linewidth}{lc@{\extracolsep{\fill}}r@{\extracolsep{0pt}}}
\ifthenelse{\boolean{pdflatex}}
{\vspace*{-1.5cm}\mbox{\!\!\!\includegraphics[width=.14\textwidth]{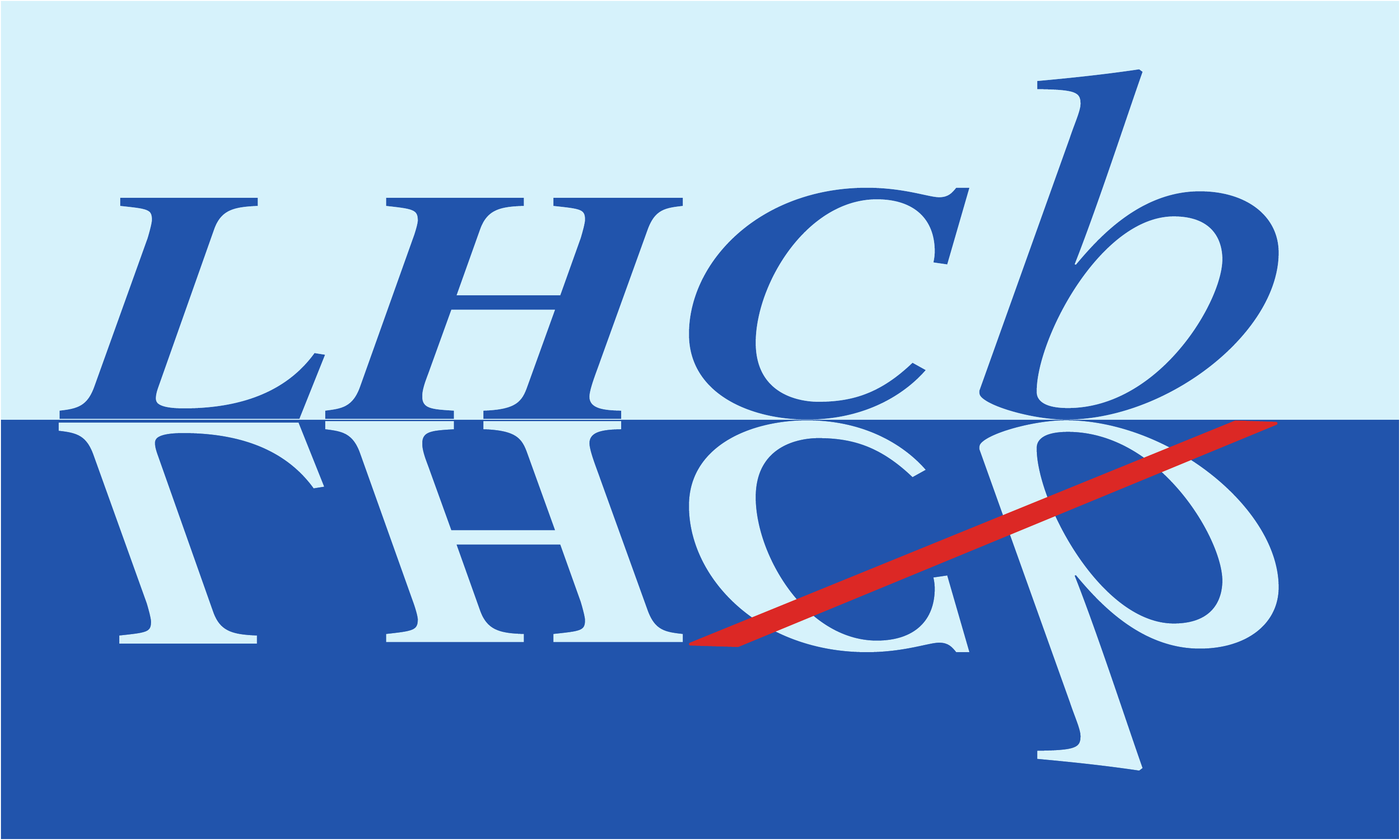}} & &}%
{\vspace*{-1.2cm}\mbox{\!\!\!\includegraphics[width=.12\textwidth]{lhcb-logo.eps}} & &}%
\\
 & & CERN-EP-2022-108 \\  
 & & LHCb-PAPER-2022-012 \\  
 & & 23 June 2023 \\ 
 & & \\
\end{tabular*}

\vspace*{4.0cm}

{\normalfont\bfseries\boldmath\huge
\begin{center}
  \papertitle 
\end{center}
}

\vspace*{1.0cm}

\begin{center}
\paperauthors 
\footnote{Authors are listed at the end of this paper.}
\end{center}

\vspace{\fill}

\begin{abstract}
  \noindent
The cross-sections of exclusive (coherent) photoproduction $J/\psi$ and $\psi(\mathrm{2S})$ mesons in ultra-peripheral PbPb collisions at a nucleon-nucleon centre-of-mass energy of $5.02\,\mathrm{TeV}$ are measured using a data sample corresponding to an integrated luminosity of $228\pm10\,\mathrm{\mu b}^{-1}$, collected by the LHCb experiment in 2018.
The differential cross-sections are measured separately as a function of transverse momentum and rapidity in the nucleus-nucleus centre-of-mass frame for $J/\psi$ and $\psi(\mathrm{2S})$ mesons. 
The integrated cross-sections are measured to be \mbox{$\sigma^\mathrm{coh}_{J/\psi} = 5.965 \pm 0.059 \pm 0.232 \pm 0.262\,\mathrm{mb}$} and
\mbox{$\sigma^\mathrm{coh}_{\psi(\mathrm{2S})} = 0.923 \pm 0.086 \pm 0.028 \pm
0.040\,\mathrm{mb}$}, 
where the first listed uncertainty is statistical, the second systematic and the third due to the luminosity determination. 
The cross-section ratio is measured to be
\mbox{$\sigma^\mathrm{coh}_{\psi(\mathrm{2S})}/\sigma^\mathrm{coh}_{J/\psi} = 0.155 \pm 0.014 \pm 0.003$}, 
where the first uncertainty is statistical and the second is systematic. 
These results are compatible with theoretical predictions.

\end{abstract}

\begin{center}
Published in JHEP 06 (2023) 146
\end{center}

\vspace{\fill}

{\footnotesize 
\centerline{\copyright~\papercopyright. \href{\paperlicenceurl}{\paperlicence}.}}
\vspace*{2mm}

\end{titlepage}


\newpage
\setcounter{page}{2}
\mbox{~}
%
%
%
%

\cleardoublepage


\renewcommand{\thefootnote}{\arabic{footnote}}
\setcounter{footnote}{0}


\cleardoublepage


\pagestyle{plain} 
\setcounter{page}{1}
\pagenumbering{arabic}


\section{Introduction}

\begin{figure}[htbp]
  \begin{center}
    \includegraphics[width=0.8\linewidth]{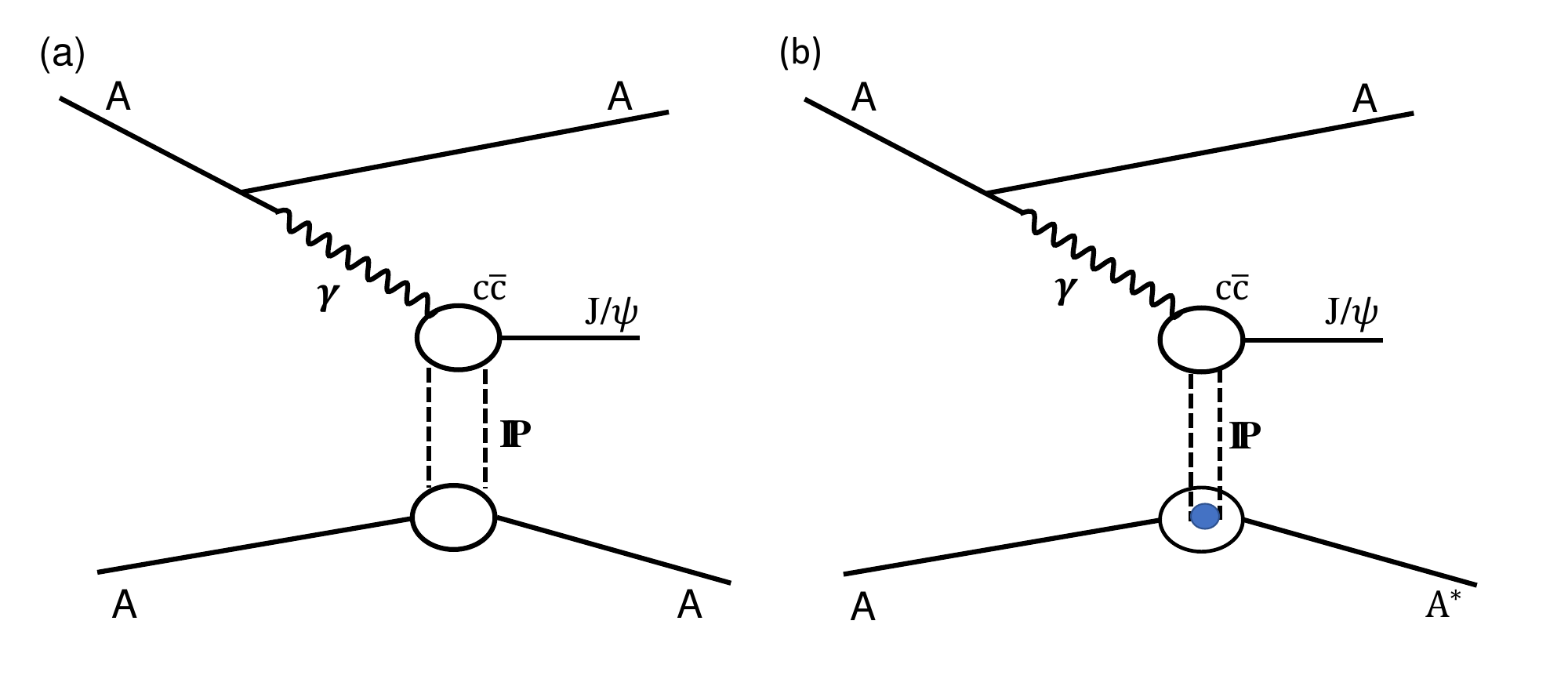}
  \end{center}
  \caption{Schematic diagrams for charmonium production in (a) coherent and (b) incoherent UPC heavy-ion collisions. For incoherent production, the pomeron is emitted from a single nucleon, indicated by the blue dot, and the nucleus typically dissociates, indicated by $A^*$. }
  \label{fig:diagram3}
\end{figure}

Ultra-peripheral collisions (UPCs) occur when two nuclei collide with an impact parameter, the distance between their centres, larger than the sum of their radii~\cite{Bertulani:2005ru}.
Because the nuclei do not overlap, strong interactions are suppressed so that photon-induced interactions between the two ions dominate.
The number of photons produced is proportional to the square of electric charge, so photon-nuclear interactions are significantly enhanced in lead-lead (PbPb) collisions compared to proton-proton ($pp$) collisions.
In UPCs, \jpsi and \psitwos mesons can be produced from the colourless exchange of a photon from one of the two nuclei and a pomeron from the other. Coherent (exclusive) photoproduction occurs when the photon couples coherently with the entire nucleus through an exchange of a pomeron, while for incoherent photoproduction, the photon interacts with a particular nucleon within the nucleus.
In this work, the terms ``coherent'' and ``incoherent'' charmonium photoproduction refer to the two diagrams, respectively, shown in Figure~\ref{fig:diagram3}.

The coherent (exclusive) photoproduction of \jpsi and
\psitwos mesons is expected to probe the nuclear
gluon
distribution functions at a momentum transfer of 
$Q^2\approx m^2/4$, where $m$ is the mass of the meson.
The photon-nuclear production of these mesons
depends on the longitudinal momentum fraction of 
gluons in the nucleus, 
$x \approx (m/\sqsnn) e^{\pm y}$, 
where $y$ is the rapidity of the meson and \sqsnn is the nucleon-nucleon centre-of-mass energy. 
Thus, coherent photoproduction of charmonium mesons provides an excellent laboratory to study nuclear shadowing effects and the initial states of collisions with small $x$, where $10^{-5} \lesssim x \lesssim 10^{-2}$ at the \lhc~\cite{Jones:2015nna}. 
The charmonia produced in this process have typical transverse momenta, \pt, smaller than $100\mevc$, with no other particles produced in the collision.
Coherent (exclusive) \jpsi photoproduction was first measured in UPCs at HERA~\cite{H1:2005dtp,ZEUS:2002wfj} in electron-proton scattering, and with ions at RHIC~\cite{PHENIX:2009xtn}.
This process has also been measured by the \cms experiment in the central region~\cite{CMS:2016itn}, by the \lhcb experiment in the forward region~\cite{LHCb:2021bfl}, and by the \alice collaboration in both the central and forward regions~\cite{ALICE:2021gpt, 2019134926} at $\sqsnn=5.02\tev$ in PbPb collisions at the \lhc.  

This paper presents a measurement of the coherent \jpsi and
\psitwos production reconstructed through the dimuon final state 
using the 2018 PbPb data sample collected by the \lhcb experiment
at $\sqsnn=5.02\tev$ and corresponding to an integrated luminosity of $228\pm10\mub^{-1}$. 
The study also measures the ratio between the 
coherent
\psitwos and \jpsi production cross-sections, 
where the uncertainties due to systematic effects 
and the luminosity determination largely cancel.
This more precise measurement will help to constrain 
theoretical predictions, where uncertainties arise from
the choice of the meson wave function in dipole 
scattering models~\cite{20171,PhysRevC.84.011902}
and the factorisation scale in perturbative QCD
models~\cite{Guzey_2016}.

The LHCb detector and simulation are described in Sec.~\ref{sec:Detector}. The selection of signal candidates and the determination of cross-sections are described in Sec.~\ref{sec:selection} and Sec.~\ref{sec:determination}, respectively. The uncertainties due to systematic effects are described in Sec.~\ref{sec:uncertainty}, while the results are presented in Sec.~\ref{sec:results} and conclusions are given in Sec.~\ref{sec:conclusion}.

\section{Detector, event reconstruction and simulation}
\label{sec:Detector}

The \lhcb detector~\cite{LHCb-DP-2008-001,LHCb-DP-2014-002} is a single-arm forward spectrometer covering the \mbox{pseudorapidity} range $2<\eta <5$, designed for the study of particles containing \bquark or \cquark quarks.
The detector includes a high-precision tracking system consisting of a silicon-strip vertex detector surrounding the collision region, a large-area silicon-strip detector located upstream of a dipole magnet with a bending power of about $4{\mathrm{\,Tm}}$, and three stations of silicon-strip detectors and straw drift tubes placed downstream of the magnet.
The tracking system provides a measurement of the momentum, \ptot, of charged particles with a relative uncertainty that varies from 0.5\% at low momentum to 1.0\% at 200\gevc.
The minimum distance of a track to a primary $pp$ collision vertex, the impact parameter (IP), is measured with a resolution of $(15+29/\pt)\mum$, where \pt is in\,\gevc.
Photons, electrons and hadrons are identified by a calorimeter system consisting of scintillating-pad (SPD) and preshower detectors, an electromagnetic and a hadronic calorimeter.
Muons are reconstructed as a long track passing through the vertex detector and the three stations of silicon-strip tracking detectors, 
and  identified by a system composed of alternating layers of iron and multiwire proportional chambers.

The pseudorapidity coverage is extended by forward shower counters (\herschel) consisting of five planes of scintillators with three planes at 114, 19.7 and $7.5\m$ upstream of the \lhcb detector, and two planes downstream at 20 and $114\m$.
The \herschel detector~\cite{Akiba_2018} significantly extends the acceptance for detecting particles from dissociated nucleons by covering the pseudorapidity range of $5 \lesssim |\eta| \lesssim 10$, enhancing the classification of central exclusive production and UPC events.

The online event selection is performed by a trigger, which consists of a hardware stage, based on information from the calorimeter and muon systems, followed by a software stage, which applies a full event reconstruction.

Simulated events are used to determine corrections for the detector resolution, acceptance, and efficiency.
The UPCs are modelled using \textsc{STARlight}~\cite{Klein:2016yzr} with a specific \lhcb configuration~\cite{LHCb-PROC-2010-056}.
The \textsc{STARlight} generator models coherent and incoherent vector-meson production in photon-nuclear interactions.
Decays of unstable particles are described by \evtgen~\cite{Lange:2001uf} with QED final-state radiation handled by \photos~\cite{davidson2015photos}.
The interactions of the generated particles with the detector are modelled using the \geant toolkit~\cite{Allison:2006ve,Agostinelli:2002hh} as described in Ref.~\cite{LHCb-PROC-2011-006}.

\section{Selection of signal candidates}
\label{sec:selection}

Signal candidates are reconstructed through the 
decays $\decay{\jpsi}\mumu$ and $\decay{\psitwos}\mumu$, and are
required to have a rapidity within the range 
$2.0 < y^* < 4.5$, where 
the starred notation indicates that 
the observable is defined in the nucleus-nucleus
centre-of-mass frame. 
All remaining selection criteria given here are
defined in the laboratory frame.
One of the candidate muons must pass the
hardware-level trigger, which requires 
a muon \pt greater than $500\mevc$. 
The dimuon candidates are selected with a
minimum-bias software trigger, 
requiring at least one 
track reconstructed by the vertex detector; 
this software trigger is $100\%$ efficient 
with respect to the following
offline selection,
since it has a looser multiplicity requirement.
The offline selection requires two muon
candidates, both with tracks that have
$\pt>700\mevc$ within the pseudorapidity 
range $2.0<\eta<4.5$. 
The dimuon candidates are required to have 
$\pt < 1 \gevc$ and an azimuthal opening
angle between the muons larger than $0.9\,\pi$. 
The mass of each signal candidate, $m_{\mumu}$, 
is required to be within $\pm 65\mevcc$ of the
known \jpsi mass~\cite{PDG2020} or 
$\pm 77.35\mevcc$ of the known \psitwos
mass~\cite{PDG2020}. 
To suppress background from 
PbPb collisions with impact parameter smaller than two times the nucleus radius, 
only events with less than 20 hits in 
the SPD are retained,
corresponding to very low occupancy events that make up about 0.3\% of all minimum-bias events.
Additionally, a requirement
based upon a figure of merit that combines 
the signals from all \herschel stations~\cite{Akiba_2018},
is used to discard events with significant
activity in the \herschel acceptance region.  
  
\section{Cross-section determination}
\label{sec:determination}

For comparison with theoretical predictions, the measured cross-sections are transformed into the nucleus-nucleus centre-of-mass frame, from the laboratory frame, 
to account for the non-zero crossing angle between the two Pb beams. 
The differential cross-section for coherent charmonium production in a given interval of rapidity or transverse momentum is determined as
\begin{equation}
\label{eq:csy}
    \frac{\deriv\sigma_{\psi}^\coh}{\deriv x}=
    \frac{N_{\psi}^\coh}{\lum\times\etot\times\BR(\decay{\psi}\mumu)\times\Delta x} \,,
\end{equation}
where $\psi$ is either \jpsi or \psitwos, 
$x$ represents either the $y^*$ or $\pt^*$, $N_{\psi}^\coh$ is the coherent signal yield, \etot is the total efficiency, \lum is the integrated luminosity, $\Delta x$ is the width of either the $y^*$- or $\pt^*$-interval, and $\BR(\decay{\psi}\mumu)$ is the branching fraction of the charmonium decay. The branching fractions \mbox{$\BR(\decay{\jpsi}\mumu)=(5.961\pm0.033)\times10^{-2}$}
and 
\mbox{$\BR(\decay{\psitwos}{\epem}) = (7.93 \pm 0.17)\times 10^{-3}$}~\cite{PDG2020} are used. 
For the \psitwos the more accurate dielectron branching fraction is used, where lepton universality is assumed.
The ratio between the differential cross-sections of \psitwos and \jpsi production in a given rapidity interval is given by
\begin{equation}
  \frac{\deriv\sigma_\psitwos^\coh/\deriv y^*}{\deriv \sigma_\jpsi^\coh/\deriv y^*} 
= \frac{N_{\psitwos}^\coh\times \varepsilon_{\jpsi}\times \BR (\decay{\jpsi} \mumu)}{N_{\jpsi}^\coh\times \varepsilon _{\psitwos}\times \BR(\decay{\psitwos} \mumu)}\,.
  \label{E3}
\end{equation}

\begin{figure}[tb]
\begin{center}
\includegraphics[width=0.80\linewidth, page={1}]{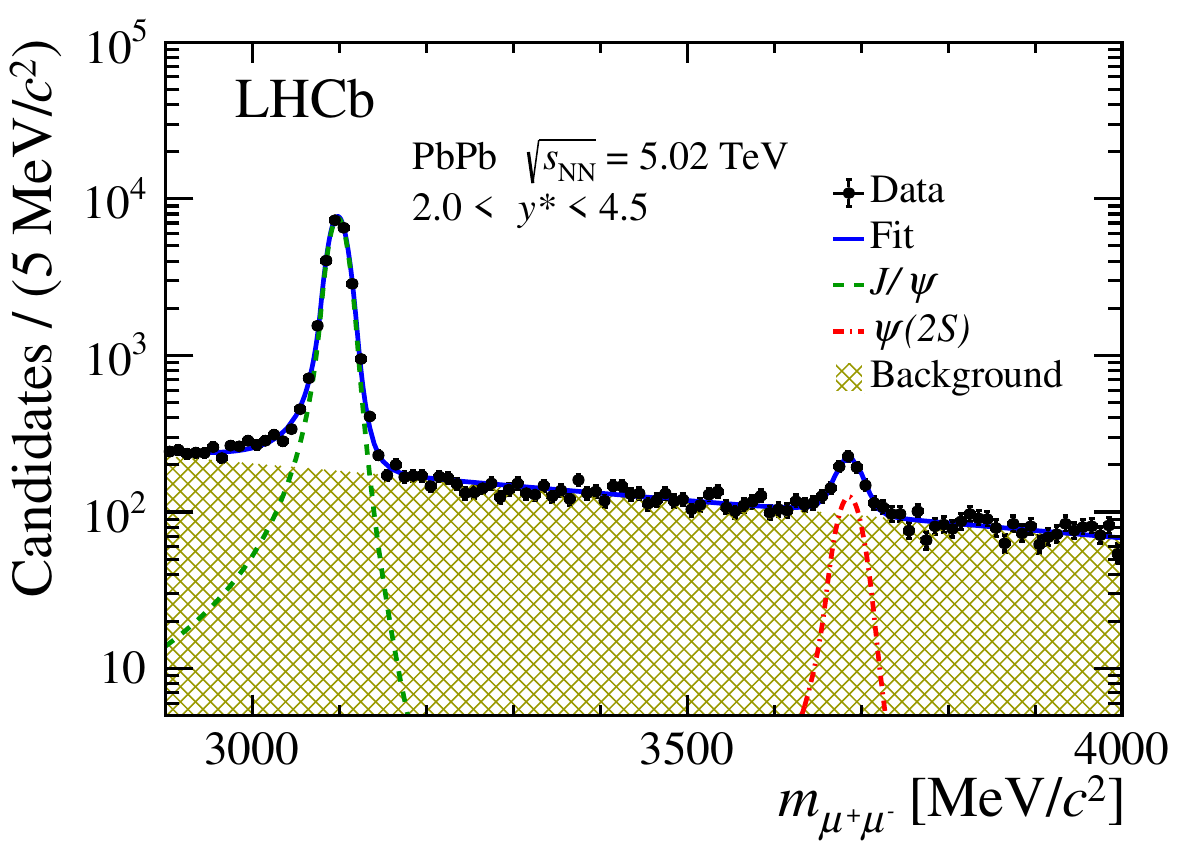}
\end{center}
\caption{Dimuon mass distribution for signal candidates in the rapidity range $2.0<y^{*}<4.5$.
The data are overlaid with the result of the fit.}
\label{fig:massfit}
\end{figure}

The signal yields are extracted in two steps.
First, an unbinned extended maximum-likelihood fit
to the dimuon mass distribution is performed to
obtain the \jpsi and \psitwos yields within the
\jpsi and \psitwos mass windows, respectively. 
The nonresonant background yield, mostly due to $\gamma\gamma\rightarrow \mumu$ process, 
is also obtained from the mass fit. 
This fit uses
double-sided Crystal-Ball functions to describe
the $\jpsi$ and $\psitwos$ mass shapes and an exponential function for the nonresonant background. 
The fit is performed in the range 
$ 2.9 < m_{\mu^{+}\mu^{-}} < 4.0 \gevcc$. 
The mass distribution and the corresponding 
fit are shown in Fig.~\ref{fig:massfit}.

In the second step, 
the coherent yields are determined with 
unbinned maximum-likelihood fits to the 
\logpt2 distributions separately for the 
candidates inside the \jpsi and \psitwos
mass windows.
The yields of \jpsi production include
contributions from coherent and incoherent production, 
and feed-down from \psitwos decays 
into \jpsi
($\decay{\psitwos}\jpsi X$). 
Similarly, the \psitwos yields include 
contributions from both coherent and 
incoherent production, while the feed-down 
contribution from higher-order charmonium excited 
states is negligible given the current
statistical precision.
The quantity \logpt2 is used because the variable \ptsq is a proxy for the typical momentum exchange, $|t|\approx \ptsq$, in an elastic scattering process, and the logarithmic distribution allows one to see the peak of the data at low exchanged momenta.
The coherent production has the smallest momentum exchange by definition, while the incoherent production gives a relatively larger transverse momentum to the \jpsi or \psitwos meson to balance the break-up of the pomeron-emitting nucleus. 
The feed-down contribution to \jpsi production also has greater transverse momentum than the coherent production to balance the other products from the \psitwos decay.
The \logpt2 shapes of coherent, incoherent and \psitwos feed-down components 
are taken from \textsc{STARlight} simulation, 
while the normalisation of these components are 
left free in the fit.
The nonresonant background consists mostly of the $\gamma\gamma \rightarrow \mumu$ process with a slightly lower transverse momentum of the dimuon system than coherent charmonium production. 
The distribution also contains a small contribution from the random pairing of uncorrelated muons produced in the hadronic interactions during peripheral or central lead-lead collisions, signified by a large transverse momentum of the dimuon system.
The \textsc{STARlight} simulation gives a precise description of \logpt2 spectrum of the $\gamma\gamma \rightarrow \mumu$ process, but not of the background from hadronic interactions.
Instead, a data-driven method is chosen to model the nonresonant background by taking the dimuon candidates in the mass range $3.2 < m_{\mumu} < 3.6 \gevcc$ outside charmonium mass windows. 
In this way, the model includes the $\gamma\gamma \rightarrow \mumu$ process and the QCD background together, and gives an unbiased modelling of the \logpt2 spectrum. 
The yields of the nonresonant background 
are determined as the integral of the nonresonant
component from the dimuon mass fit separately in
the \jpsi and \psitwos mass windows, and are fixed in the \logpt2 fits.

\begin{figure}[tb]
    \centering
    \hfil
    \begin{minipage}[t]{0.49\linewidth}
        \centering
        \includegraphics[width=\linewidth, page = {1}]{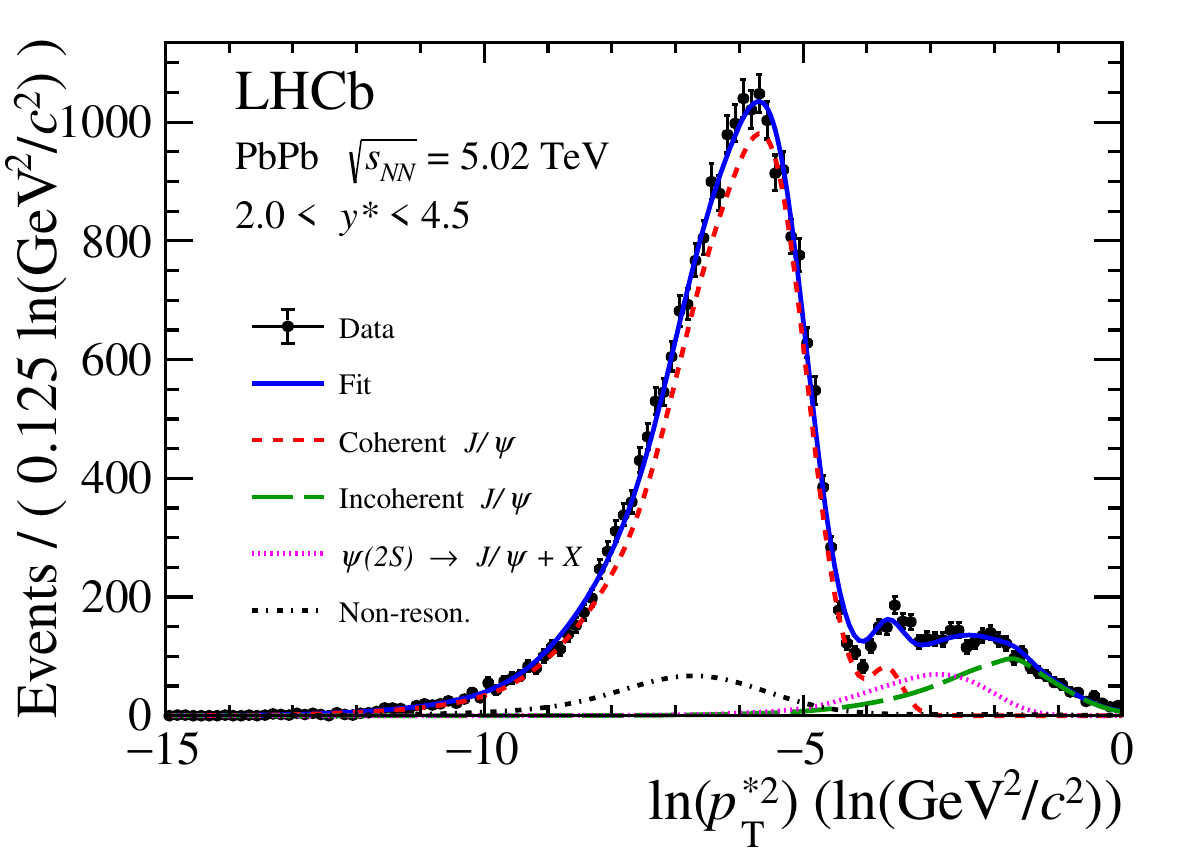}
        \put(-50,130){\jpsi}
    \end{minipage}
    \begin{minipage}[t]{0.49\linewidth}
        \centering
        \includegraphics[width=\linewidth, page = {1}]{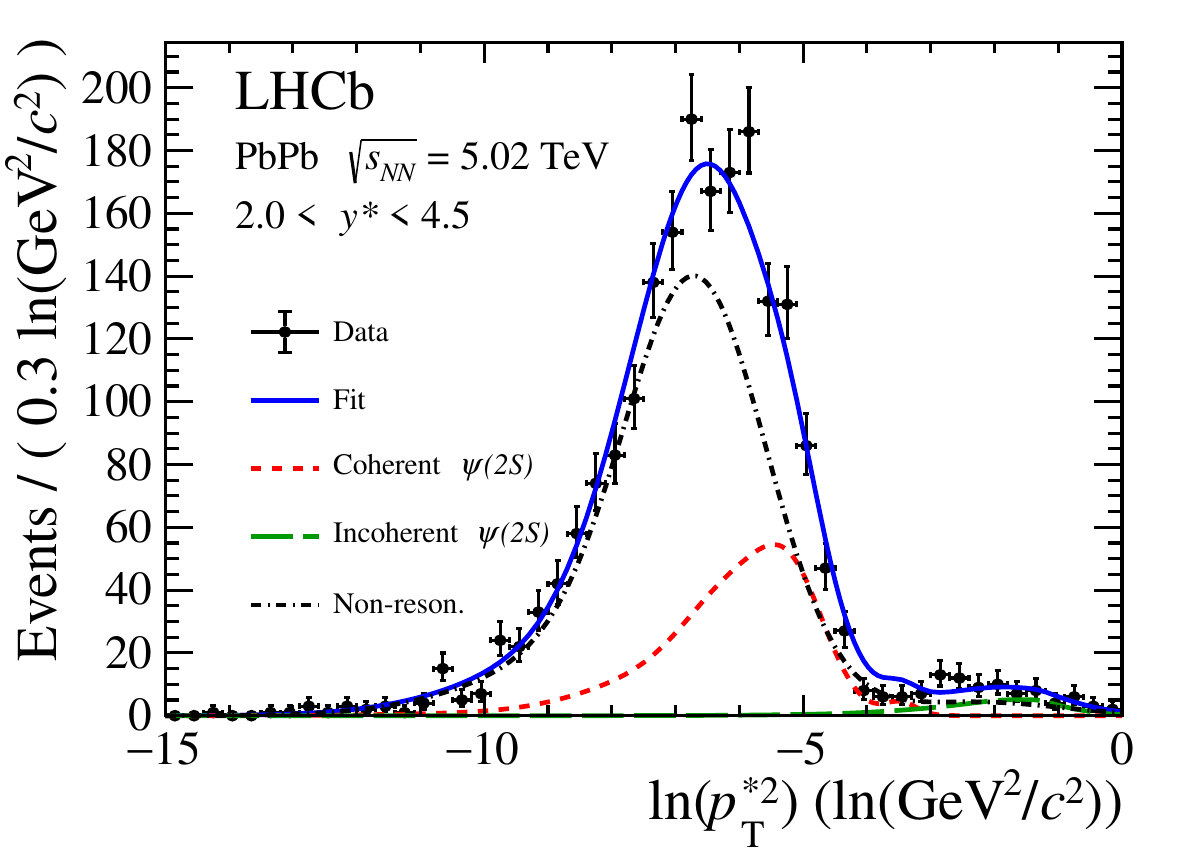}
        \put(-50,130){\psitwos}
    \end{minipage}
    \hfil
    \caption{The \logpt2 distribution of dimuon candidates in the $2.0<y^{*}<4.5$ range
    for (left) \jpsi candidates and (right) 
    \psitwos candidates. 
    The data are overlaid with the result of the fit.}
    \label{fig:2d}
\end{figure}

Figure~\ref{fig:2d} shows the \logpt2 distributions 
of selected \jpsi and \psitwos candidates in the
rapidity interval $2 < y^* < 4.5$. 
Fits to the \logpt2 distributions are performed 
in each $y^*$ interval to extract the corresponding 
\jpsi and \psitwos yields, 
as reported 
in Table~\ref{tab:MassFitResult}.
The coherent yield of 
\jpsi and \psitwos 
production
for each $\pt^*$ interval
is calculated by subtracting the background
components from the measured 
yield for that interval
as reported in
Tables~\ref{tab:yields_jpsi_events_pt} and
\ref{tab:yields_psi2s_events_pt}.
The contributions from background components are 
determined by an overall fit to the 
\logpt2 distributions.

\begin{table}[htbp]
    \caption{Total and coherent \jpsi and \psitwos yields from the invariant mass and transverse momentum fits in different rapidity intervals.}
    \def\pz{\phantom{0}}
    \label{tab:MassFitResult}
      \begin{center}
      \begin{tabular}{crrrr}
      \hline
      Interval & \multicolumn{1}{c}{$N^\tot_{\jpsi}$} & \multicolumn{1}{c}{$N^\coh_{\jpsi}$} & \multicolumn{1}{c}{$N^\tot_{\psitwos}$} & \multicolumn{1}{c}{$N^\coh_{\psitwos}$}\\
      \hline
$2.0<y^*<4.5$ &$ 23\,355\pm 183$ &$ 20\,193\pm 199$ &$ 513\pm 43$ &$ 471\pm 44$ \\
\hline
$2.0<y^*<2.5$ &$ 2\,457\pm\pz 60$ &$ 2\,070\pm \pz66$ &$ 75\pm 15$ &$ 65\pm 15$ \\
$2.5<y^*<3.0$ &$ 6\,845\pm 100$ &$ 5\,926\pm 108$ &$ 147\pm 26$ &$ 137\pm 26$ \\
$3.0<y^*<3.5$ &$ 7\,875\pm 106$ &$ 6\,883\pm 115$ &$ 168\pm 26$ &$ 161\pm 26$ \\
$3.5<y^*<4.0$ &$ 5\,019\pm \pz82$ &$ 4\,362\pm \pz90$ &$ 102\pm 18$ &$ 85\pm 18$ \\
$4.0<y^*<4.5$ &$ 1\,166\pm \pz38$ &$ 956\pm \pz44$ &$ 24\pm \pz8$ &$ 21\pm\pz 8$ \\

      \hline
 \end{tabular}
    \end{center}
   \end{table} 

The total efficiency \etot is determined as the product of the acceptance efficiency ($\varepsilon_{\text{acc}}$), the muon acceptance efficiency ($\varepsilon_{\mu\text{-acc}}$), the tracking efficiency ($\varepsilon_{\text{trk}}$), the selection efficiency ($\varepsilon_{\text{sel}}$), the particle identification (PID) efficiency ($\varepsilon_{\text{PID}}$), the trigger efficiency ($\varepsilon_{\text{trg}}$) and the \herschel-veto efficiency ($\varepsilon_{\text{her}}$). 
Each efficiency is evaluated separately for \jpsi and \psitwos mesons in each $y^*$ and $\pt^*$ interval for the differential cross-section measurements.
Efficiencies are evaluated from simulation calibrated to data.
The value of $\varepsilon_{\mu\text{-acc}}$ is determined at generator level as the fraction of events with both muon candidates passing $\pt>700\mevc$ and $2.0<\eta<4.5$. The signal candidates are required to pass the $\pt<1\gevc$ selection and fall in the mass windows defined in Sec.~\ref{sec:selection}, for \jpsi and \psitwos mesons separately.
 For $\varepsilon_{\text{trk}}$, $\varepsilon_{\text{PID}}$ and $\varepsilon_{\text{trg}}$, the simulation does not always describe the data well. Efficiency corrections from data using the tag-and-probe method~\cite{LHCb-DP-2013-002} are determined from \jpsi events in PbPb collision data.
 The \herschel-veto criteria is chosen to retain 
 a signal efficiency of 90\% according to a set of 
 separately selected pure signal and background data samples.
Dependencies of the efficiency correction on $y^*$ and $\pt^*$ of the dimuon system are studied and found to be negligible in the invariant mass range from $2.9$ to $4.0\gevcc$.

   \begin{table}[htbp]
   \def\pz{\phantom{0}}
     \caption{Total and coherent \jpsi yields in different $\pt^*$ intervals within the rapidity range $2.0<y^*<4.5$.}
     \label{tab:yields_jpsi_events_pt}
       \begin{center}
       \begin{tabular}{rrr}
       \hline
       Interval \unitmevc  & \multicolumn{1}{c}{$N_{\jpsi}^{\tot}$} &  \multicolumn{1}{c}{$N_{\jpsi}^{\coh}$}\\
      \hline
$0<\pt^*<200$ &$ 21\,153\pm 175$ &$ 20\,180\pm 175$ \\
\hline
$0<\pt^*<\pz20$ &$ 2\,216\pm\pz 58$ &$ 2\,204\pm\pz 58$ \\
$20<\pt^*<\pz40$ &$ 5\,647\pm\pz 92$ &$ 5\,619\pm\pz 92$ \\
$40<\pt^*<\pz60$ &$ 5\,931\pm\pz 83$ &$ 5\,885\pm\pz 83$ \\
$60<\pt^*<\pz80$ &$ 3\,928\pm\pz 65$ &$ 3\,863\pm\pz 65$ \\
$80<\pt^*<100$ &$ 1\,848\pm\pz 44$ &$ 1\,759\pm\pz 44$ \\
$100<\pt^*<120$ &$ 497\pm\pz 23$ &$ 381\pm\pz 24$ \\
$120<\pt^*<140$ &$ 225\pm\pz 16$ &$ 88\pm\pz 17$ \\
$140<\pt^*<160$ &$ 289\pm\pz 17$ &$ 137\pm\pz 18$ \\
$160<\pt^*<180$ &$ 328\pm\pz 18$ &$ 167\pm\pz 20$ \\
$180<\pt^*<200$ &$ 244\pm\pz 16$ &$ 77\pm\pz 17$ \\
      \hline
      \end{tabular}
      \end{center}
   \end{table}

   \begin{table}[htbp]
   \def\pz{\phantom{0}}
     \caption{Total and coherent \psitwos yields in different $\pt^*$ intervals within the rapidity range $2.0<y^*<4.5$.}
     \label{tab:yields_psi2s_events_pt}
       \begin{center}
       \begin{tabular}{rrr}
       \hline
       Interval \unitmevc & \multicolumn{1}{c}{$N_{\psitwos}^{\tot}$} & \multicolumn{1}{c}{$N_{\psitwos}^{\coh}$}\\
      \hline
$0<\pt^*<200$ &$ 475\pm 41$ &$ 468\pm 41$ \\
\hline
$0<\pt^*<\pz30$ &$ 77\pm 35$ &$ 77\pm 35$ \\
$30<\pt^*<\pz70$ &$ 275\pm 39$ &$ 274\pm 39$ \\
$70<\pt^*<\pz90$ &$ 91\pm 14$ &$ 91\pm 14$ \\
$90<\pt^*<110$ &$ 27\pm\pz 8$ &$ 27\pm\pz 8$ \\
$110<\pt^*<150$ &$ 0\pm\pz 5$ &$ 0\pm\pz 5$ \\
$150<\pt^*<200$ &$ 5\pm\pz 4$ &$ 2\pm\pz 4$ \\

      \hline
      \end{tabular}
      \end{center}
   \end{table}

\section{Systematic uncertainties}
\label{sec:uncertainty}

Systematic uncertainties on the 
cross-section measurements arise from the
efficiency and background determination, 
signal and background shapes, 
momentum resolution, 
integrated luminosity and 
knowledge of the $\decay{\jpsi}\mumu$ and
$\decay{\psitwos}\mumu$ branching fractions.
For the \psitwos to \jpsi cross-section 
ratio measurement, 
only systematic uncertainties from the 
charmonia decay branching fractions 
are considered.
Those from efficiency and background
determination, 
signal and background shapes
integrated luminosity 
are highly correlated and cancel. 
A summary of the systematic uncertainties is presented in Table~\ref{tab:summary}.

The systematic uncertainties related to the
efficiencies are driven by the sizes of 
the simulation and data samples.
They vary from (0.5\,--\,2.0)\% 
for the tracking efficiency,
(0.9\,--\,1.6)\% for the PID efficiency and
(2.1\,--\,3.7)\% for the trigger efficiency,
across different $y^*$ and $\pt^*$ intervals.
The uncertainty associated with the \herschel
efficiency is a constant 1.4\%.

The uncertainty on the background shape is
estimated by varying the shape parameters 
within their fitted uncertainties. 
The maximum difference on the extracted
signal yields is 1.2\%, and is assigned
as the background uncertainty.

The momentum resolution is expected to 
shift events from one $\pt^*$ interval 
to another. 
The uncertainties due to the momentum 
resolution are evaluated by comparing 
the \pt spectra between 
generated and reconstructed events. 
The evaluated relative uncertainties 
vary from 0.9 to 34\% 
for different $\pt^*$ intervals.
The largest uncertainty 
corresponds to the $\pt^*$ interval 
between 140 to 160\mev 
as shown in Table~\ref{tab:jpsipt_cross_245} 
(Appendix~\ref{sec:A}),
where very small signal yields
are observed.

The slight discrepancies between the data 
and the fit results are visible at 
\mbox{$\logpt2 \sim -4$\,[$\ln{(\gevc^2)}$]} 
for both the \jpsi and \psitwos fits, 
as seen in Fig.~\ref{fig:2d}.
This is expected to originate from a mis-modelling 
of the predicted signal shape from simulation.
A systematic uncertainty on the signal shape 
model is estimated by evaluating the difference
between the fitted signal yields 
with respect to an alternative 
empirical signal shape.
The obtained difference is about 0.04\%,
negligible compared to other uncertainties shown in Table~\ref{tab:summary}.

The uncertainties on the branching fractions result in relative uncertainties on the measured cross-sections of 0.6\% and 2.1\%~\cite{PDG2020}, respectively. The relative uncertainty on the luminosity is 4.4\%~\cite{LHCb-PAPER-2014-047}.

\begin{table}[htbp]
    \caption{Summary of the systematic uncertainties.}
  \label{tab:summary}
\begin{center}
\begin{tabular}{lcc}
\hline
 \multicolumn{1}{c}{Source} 
& \multicolumn{2}{c}{Relative uncertainty [\%]} \\
\cline{2-3}
& $\sigma_\jpsi^\coh$ & $\sigma_\psitwos^\coh$ \\
\hline
Tracking efficiency & 0.5--2.0 & 0.5--2.0 \\
PID efficiency       & 0.9--1.6 & 0.9--1.6 \\
Trigger efficiency       & 2.7--3.7 & 2.1--2.5 \\
$\herschel$ efficiency & 1.4 & 1.4 \\
Background estimation   & 1.2    &  1.2 \\
Momentum resolution  & 0.9--34 & 1.3--27 \\
Branching fraction   & 0.6 & 2.1 \\
Luminosity    & 4.4 & 4.4 \\
\hline

\end{tabular}
\end{center}
\end{table}

\section{Results and discussion}
\label{sec:results}

The integrated cross-sections of 
coherent $\jpsi$ and $\psitwos$ 
photoproduction in PbPb collisions are measured 
in the rapidity region $2.0 < y^* < 4.5$ as
\begin{align*}
\sigma^\coh_{\jpsi} &= 5.965 \pm 0.059 \pm 0.232 \pm 0.262 \mbarn\,, \\
\sigma^\coh_{\psitwos} &= 0.923 \pm 0.086 \pm 0.028 \pm 0.040 \mbarn\,, 
\end{align*}
where the first listed uncertainty
is statistical, 
the second is systematic
and the third is due to
the luminosity determination.
The cross-section ratio between coherent 
\psitwos and \jpsi photoproduction is measured to be
\begin{align*}
\sigma^\coh_{\psitwos}/\sigma^\coh_{\jpsi} &= 0.155 \pm 0.014 \pm 0.003\,,
\end{align*}
where the first uncertainty is 
statistical and the second is systematic. 
The luminosity uncertainty cancels 
in the ratio measurement.

\begin{figure}[htbp]
    \centering
    \begin{minipage}[t]{0.423\linewidth}
        \centering
        \includegraphics[trim=0 0 180 0,clip,width=\linewidth]{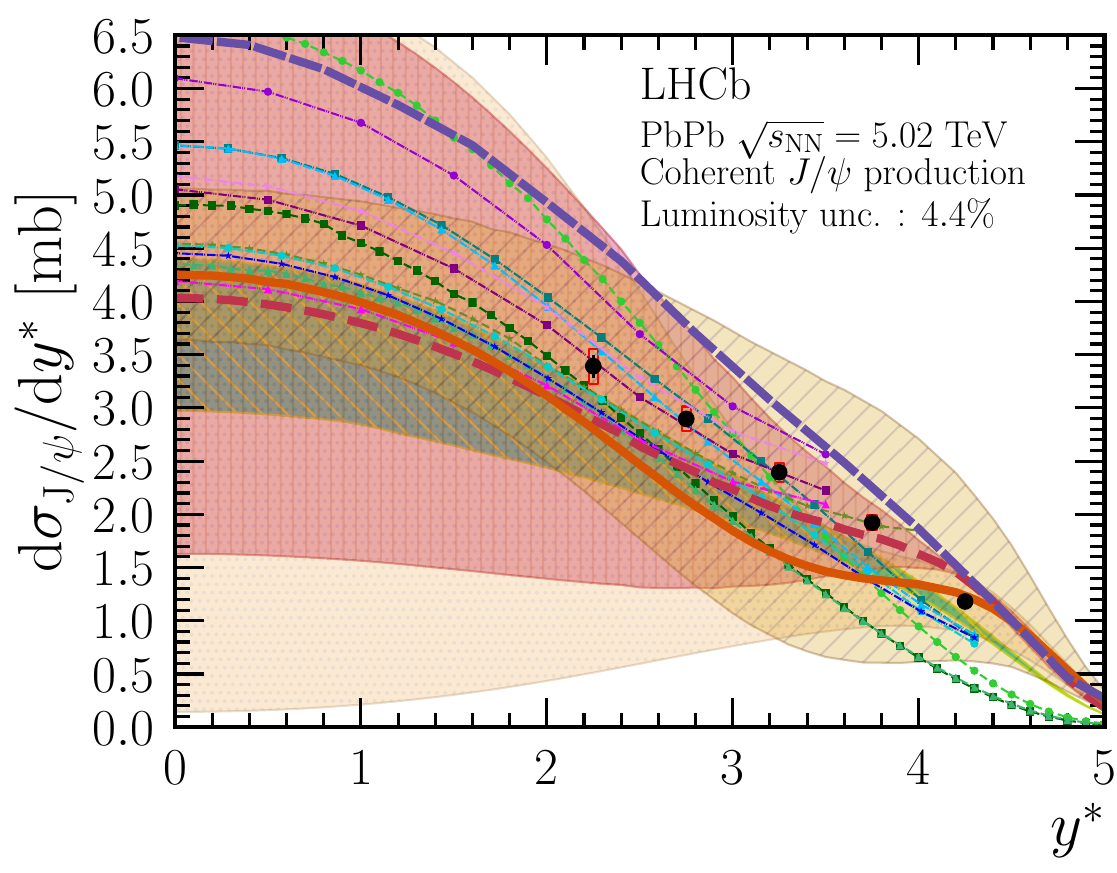}
    \end{minipage}
    \begin{minipage}[t]{0.567\linewidth}
        \centering
        \includegraphics[width=\linewidth]{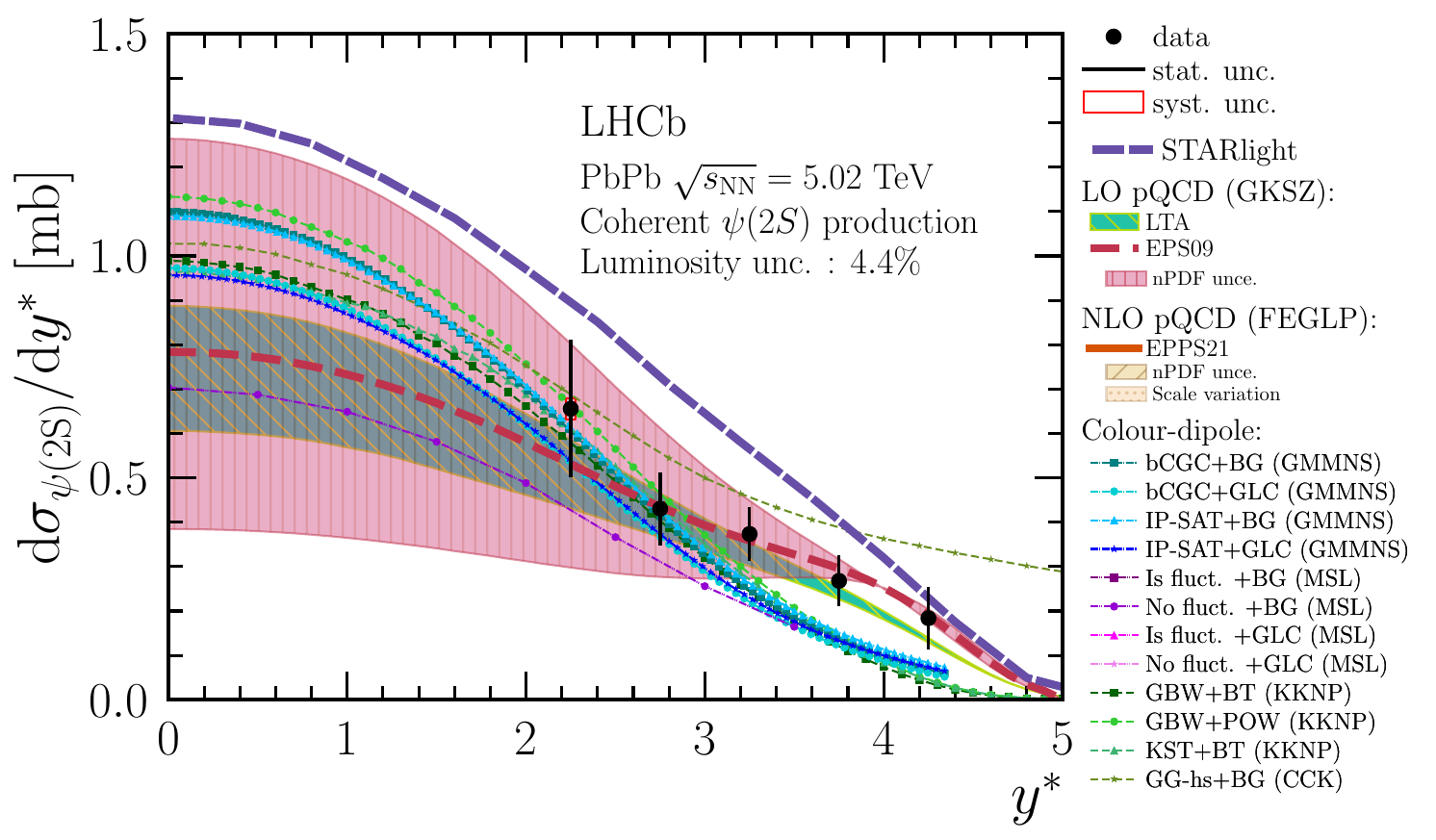}
     \end{minipage}
    \vspace{-20pt}
    \caption{Differential cross-section as a function $y^*$ for coherent (left) $\jpsi$ and (right) \psitwos photoproduction, compared to theoretical predictions. }
    \label{fig:theo_y}
\end{figure}

\begin{figure}[htbp]
    \centering
    \begin{minipage}[t]{0.425\linewidth}
        \centering
        \includegraphics[trim=0 0 166 0,clip,width=\linewidth]{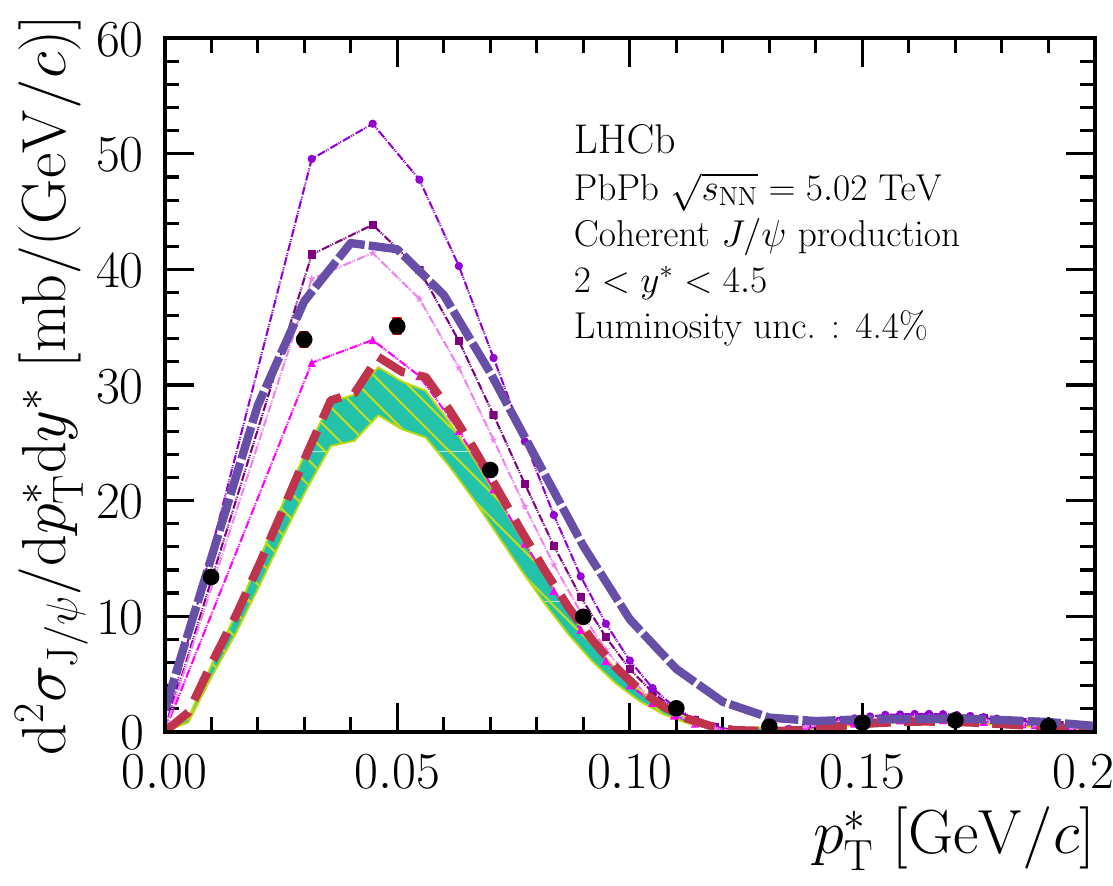}
    \end{minipage}
    \begin{minipage}[t]{0.565\linewidth}
        \centering
        \includegraphics[width=\linewidth]{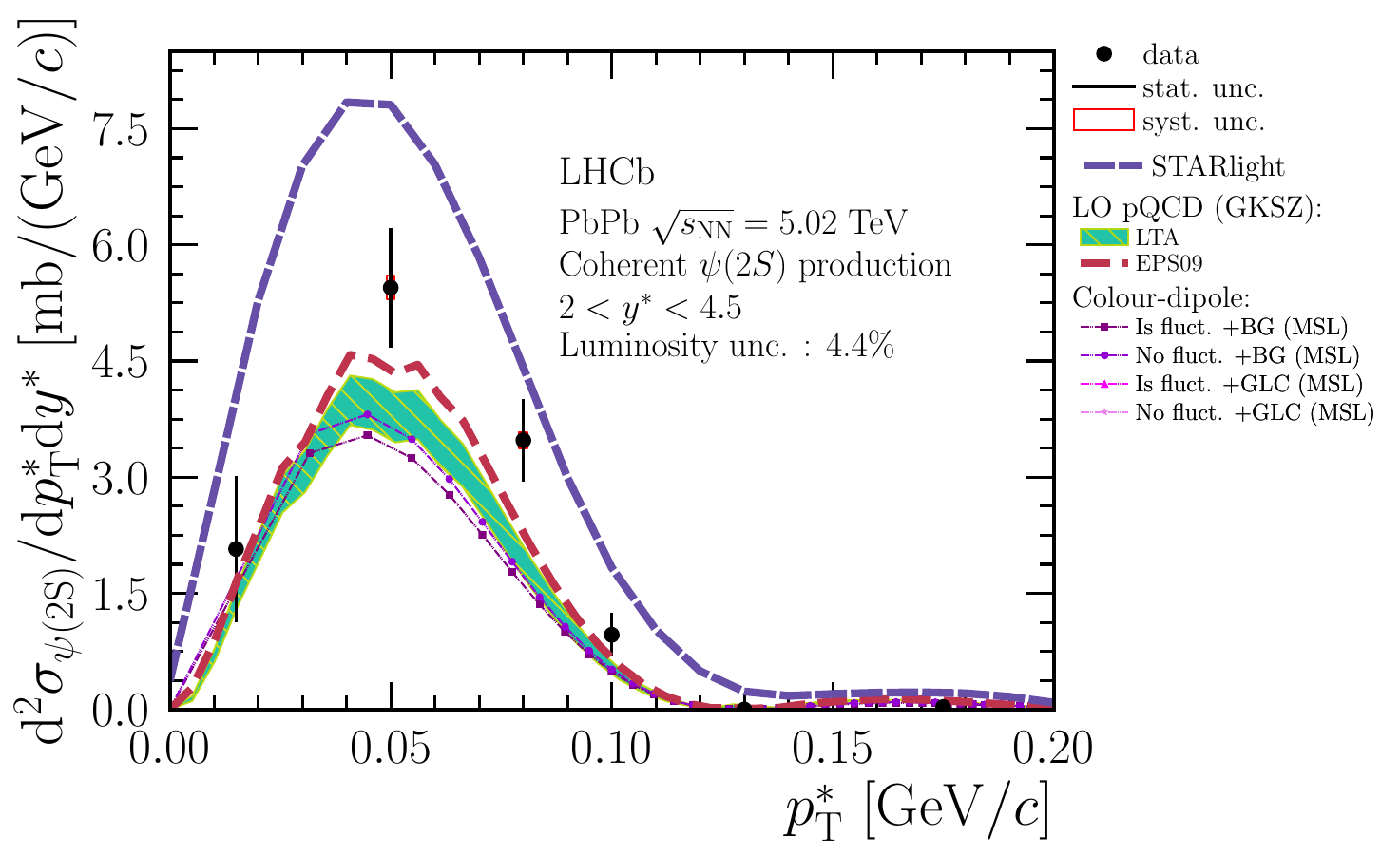}
    \end{minipage}
    \vspace{-20pt}
    \caption{Differential cross-section as a function of $\pt^*$ within the rapidity range $2<y^*<4.5$ for coherent (left) $\jpsi$ and (right) \psitwos photoproduction compared to theoretical predictions.}
    \label{fig:theo_pt}
\end{figure}

\begin{figure}[htbp]
\begin{center}
\includegraphics[width=0.88\linewidth]{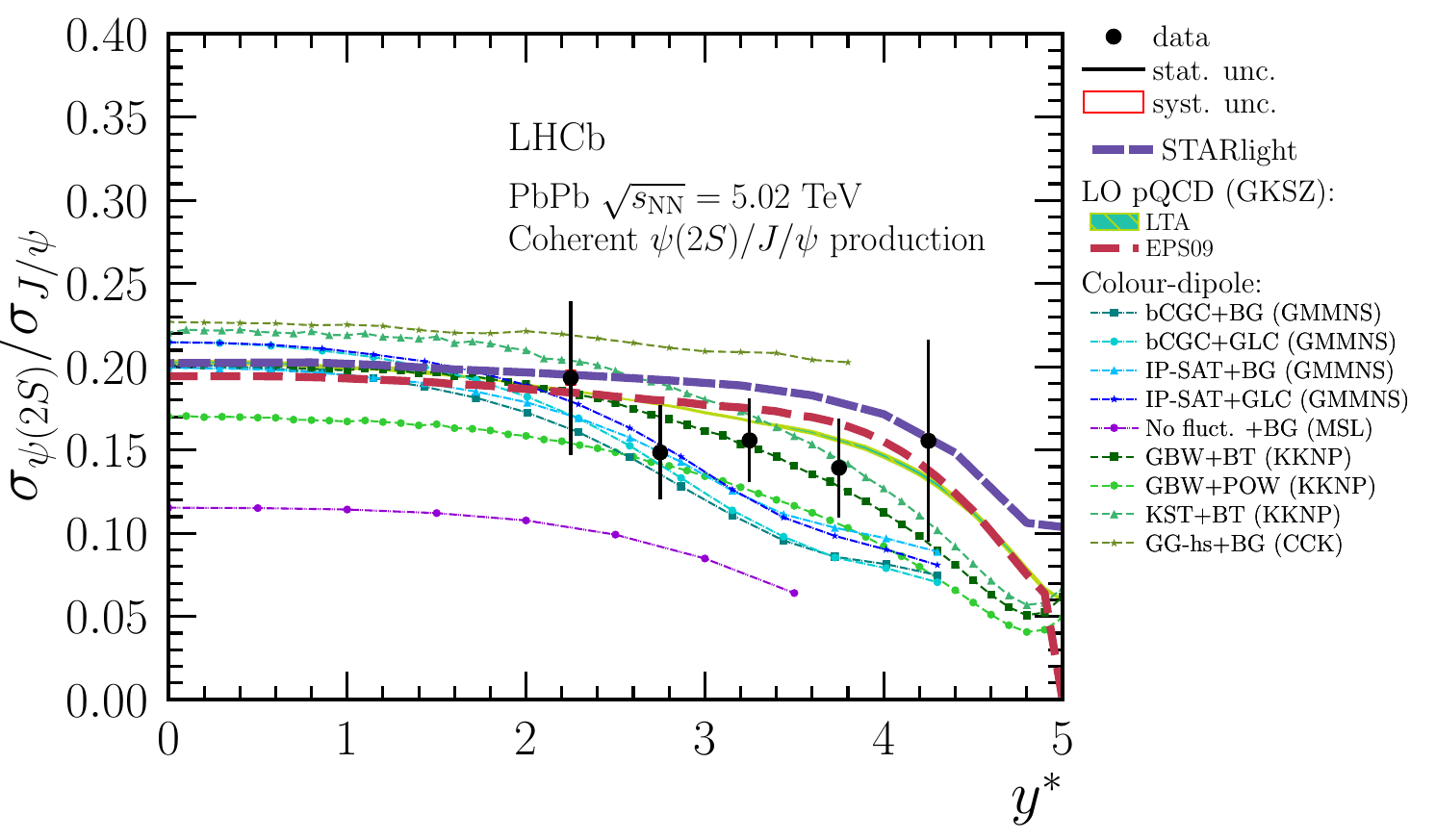}
\end{center}
\vspace{-20pt}
\caption{Differential cross-section ratio of \psitwos to \jpsi coherent photoproduction as a function of $y^*$, compared to theoretical predictions.}
\label{fig:y_theo_ratio}
\end{figure}

The measured differential cross-sections for coherent \jpsi and \psitwos photoproduction as functions of $y^*$ and $\pt^*$ are shown in Figs.~\ref{fig:theo_y} and \ref{fig:theo_pt}, respectively.
The cross-section ratio of coherent photoproduction between \psitwos and \jpsi as a function of rapidity is shown in Fig.~\ref{fig:y_theo_ratio}. 
The data are shown as black points with black error bars for the statistical uncertainties, red boxes show the systematic uncertainties and the fully correlated uncertainty due to integrated luminosity is labelled separately.
In the same figures, the results are compared to several theoretical
predictions.
The numerical values of the results are reported in Tables~\ref{tab:jpsi_cross}\,--\,\ref{tab:psi2spt_cross_245} in Appendix~\ref{sec:A}.

The \textsc{STARlight} prediction is based on the concept of vector meson dominance with parameters tuned according to previous UPC data~\cite{Klein:2016yzr}. As shown in Figs.~\ref{fig:theo_y} and \ref{fig:theo_pt}, it gives a good description of the decreasing slope as a function of $y^*$ and the shape as a function of $\pt^*$, but the overall predicted normalisation is about $20\%$ and $50\%$ higher for \jpsi and \psitwos production, respectively.
The ratio between \psitwos and \jpsi production in Fig.~\ref{fig:y_theo_ratio} is also well modelled within data uncertainties.

Two sets of calculations using leading-order perturbative QCD (LO pQCD) are provided by Guzey, Kryshen, Strikman and Zhalov~\cite{Guzey_2016,2017access} (GKSZ) for both \jpsi and \psitwos coherent photoproduction. 
One uses the leading twist approximation (LTA)~\cite{Frankfurt:2011cs} to model the nuclear shadowing effect in the initial state.
The shaded area labelled ``LTA'' in Fig.~\ref{fig:theo_y} corresponds to the uncertainties on the nuclear shadowing determined in Ref.~\cite{Frankfurt:2011cs}. 
The other uses EPS09 nuclear parton distribution functions (nPDFs)~\cite{Eskola:2009uj} for the nuclear shadowing, with an error band labelled ``nPDF unce.'' under ``EPS09'' in Fig.~\ref{fig:theo_y} presenting the uncertainties of the nuclear modification.
Note that the two LO pQCD calculations carry {\it ad hoc} normalisation factors of the cross-section determined using high-energy HERA data~\cite{ZEUS:1996esk, Guzey_2016}. 
Both of them predict well the shapes of the data for both \jpsi and \psitwos production as a function of $y^*$ in Fig.~\ref{fig:theo_y}. 
A slightly larger (smaller) $\pt^*$ is predicted for \jpsi (\psitwos) production than the data in Fig.~\ref{fig:theo_pt}. 
An underestimation of about $15\%$  of the normalisation can be seen for both \jpsi and \psitwos production, but the ratio is well modelled in Fig.~\ref{fig:y_theo_ratio}.
The large nPDF uncertainties in Fig.~\ref{fig:theo_y} indicate that coherent charmonium photoproduction in heavy ion collisions is very sensitive to the nuclear modification factors, especially to the modelling of the gluon shadowing, used in the LO pQCD calculations~\cite{Guzey_2016}.  

The next-to-leading-order (NLO) pQCD calculation using the most recent EPPS21 NLO nPDFs~\cite{Eskola:2021nhw} is provided by Flett, Eskola, Guzey, L\"oyt\"ainen and Paukkunen~\cite{Eskola:2022vpi} (FEGLP), and is only available for \jpsi production as shown in the left plot of Fig.~\ref{fig:theo_y}. 
This is the first pQCD calculation without using {\it ad hoc} normalisation factors of the cross-section compared to previous LO calculations.
The predicted central value is about $15-20\%$ lower than the data, which is calculated based on a factorization/renormalisation scale, $\mu=0.76\,m_{\jpsi}= 2.37\gev$, tuned using previous \alice~\cite{ALICE:2012yye,ALICE:2013wjo,ALICE:2021gpt,2019134926}, \lhcb~\cite{LHCb:2021bfl} and \cms~\cite{CMS:2016itn} data. 
The substantial shaded area labelled ``scale variation'' corresponds to a variation of $\mu$ from $m_{\jpsi}/2$ to $m_{\jpsi}$, indicating that the cross-section is extremely sensitive to the missing higher-order pQCD corrections.
The nPDF uncertainties are much smaller for rapidity below 2 but much bigger for rapidity greater than 3 in the NLO pQCD calculation compared to the LO calculation. This is understood as an interplay of the real and imaginary parts of the quark and gluon amplitudes that causes a certain level of mutual cancellation of the nuclear effects, especially at lower rapidity region~\cite{Eskola:2022vpi}. High-precision data can nevertheless help to further understand these effects. 

Various calculations within the framework of the colour-dipole model ~\cite{PhysRevC.84.011902,2018,
Kopeliovich:2020has,PhysRevD.96.094027,
Gon_alves_2005,Mantysaari:2017dwh,
2014arXiv1406.2877L} are also compared in Fig.~\ref{fig:theo_y} for both \jpsi (left) and \psitwos (right). 
All these models inherit parameters determined by previous HERA data~\cite{ZEUS:2002wfj,H1:2005dtp,ZEUS:1996esk}.

The calculations provided by Gon\c{c}alves, Machado, Moreira, Navarra, Sampaio dos Santos~\cite{PhysRevD.96.094027,Gon_alves_2005} (GMMNS) employ the impact-parameter-color-gluon-condensate (bCGC)~\cite{Kowalski:2006hc,Watt:2007nr}
and the impact-parameter-saturation (IP-SAT)~\cite{Bartels:2002cj} parameterisations for the dipole-proton cross-section, combined with boosted Gaussian (BG) and Gaussian-light-cone (GLC) models~\cite{PhysRevD.96.094027} of the vector-meson wave functions. 
They agree with the \jpsi data reasonably well for rapidity below 3,
while an underestimation of about 10--30\% can be observed in Figs.~\ref{fig:theo_y} and \ref{fig:theo_pt} for rapidity above 3.5 in the case of \jpsi, and about 20--60\% for rapidity above 3 for \psitwos. 
This results in a systematic suppression of the ratio between \psitwos and \jpsi for rapidity above 3 as seen in Fig.~\ref{fig:y_theo_ratio}.

Predictions calculated by M\"antysaari, Schenke and Lappi~\cite{2014arXiv1406.2877L,Mantysaari:2017dwh} (MSL) use the IP-SAT parameterisation to describe the dipole-proton cross-section
but include sub-nucleon scale fluctuations. 
Calculations with (Is fluct.) and without (No fluct.) sub-nucleon fluctuation together with BG or GLC vector-meson wave functions are compared to the measurements in Figs.~\ref{fig:theo_y} and \ref{fig:theo_pt}.
Only the No fluct.+BG calculation is available for \psitwos as a function of $y^*$ and only No fluct.+BG and Is fluct.+BG are available for \psitwos as a function of $\pt^*$.
They predict well the shapes of the \jpsi differential cross-section as functions of $y^*$ and $\pt^*$, but the predicted $\pt^*$ for \psitwos production is slightly smaller than the data. 
Variations in the normalisation are relatively large among these models.
The two models using BG vector-meson wave functions predict higher normalisation than the two using GLC. 
Because these models are calculated for rapidity below 3.5, the normalisation of the predictions as a function of $\pt^*$ appears relatively lower than that as a function of $y^*$. 
Calculations for \psitwos are relatively worse than for \jpsi because we know less about the \psitwos wave function than \jpsi~\cite{20171,2014arXiv1406.2877L,Kowalski:2006hc}, the precise data can nevertheless be helpful to improve this aspect.
Among them, Is fluct.+BG gives the best prediction for \jpsi as a function of $y^*$.

Models by Kopeliovich, Krelina, Nemchik and Potashnikova~\cite{Kopeliovich:2020has} (KKNP) are composed of quarkonium wave functions determined by the Buchm\"uller--Tye (BT)~\cite{Buchmuller:1980su} or power-like (POW)~\cite{Martin:1980jx,Barik:1980ai} potentials, as well as the Golec--Biernat--Wusthoff (GBW)~\cite{Golec-Biernat:1998zce,Golec-Biernat:1999qor} or Kopeliovich--Schafer--Tarasov (KST)~\cite{Kopeliovich:1999am} models for the dipole-nucleon cross-sections. 
They appear similar to the models provided by GMMNS, with reasonably good agreement for \jpsi production for $y^* <3$, but with an overestimation of the decreasing slope, and consequently an underestimation of the data of 20--60\%  for $y^* > 3$ for \psitwos production.
But the decreasing slopes are consistent between the \jpsi and \psitwos calculations so that the predicted ratio between the two agrees well with the data. 

In an alternative approach (GG-hs+BG) by Cepila, Contreras and Krelina~\cite{2018} (CCK), the BG model is used for the vector-meson wave function, and the dipole-nucleon cross-section is parameterised assuming the nucleon is composed of so-called hot-spots (hs), regions with high-gluon density. The standard Glauber-Gribov (GG) formalism~\cite{Gribov:1968jf,Hufner:1996dr,Kopeliovich:2016jjx} is then used to extend the dipole-nucleon cross-section to the case for dipole-nucleus. 
This model describes well the slope as a function of $y^*$ for both \jpsi and \psitwos data, but a relatively large overestimation of the normalisation for \psitwos production. 
The corresponding prediction for the ratio between \psitwos and \jpsi production is therefore relatively higher than the data points. 

In a closer look at the differential cross-section as a function of rapidity in Fig.~\ref{fig:theo_y} for both \jpsi and \psitwos mesons, one can observe that the data do not consistently decrease at a fixed slope, instead it has a subtle and elusive bump between 3 and 4. 
This is the first observation of this subtle signature thanks to the high precision data. 
Among the models discussed above, only the standard pQCD calculations can reproduce this feature, and can be understood as an interplay of the real and imaginary parts of the quark and gluon amplitudes~\cite{Eskola:2022vpi,Guzey_2016}.

\section{Conclusion}
\label{sec:conclusion}
The coherent (exclusive) \jpsi and \psitwos photoproduction cross-sections in PbPb ultra-peripheral collisions at a centre-of-mass energy of $\sqsnn=5.02\tev$ are studied using a data sample corresponding to an integrated luminosity of $228\pm10\mub^{-1}$
collected by the \lhcb detector.
The differential cross-sections, as a function of $y^*$ and $\pt^*$, are measured separately for \jpsi and \psitwos mesons in the ranges $2.0<y^*<4.5$ and $0<\pt^*<0.2\gevc$.
The ratio of the cross-sections between the coherent \psitwos and \jpsi production, as a function of rapidity, is also determined for the first time in PbPb collisions and is found to be 
compatible with theoretical models.
The \jpsi results are the most precise measurement to date, while the \psitwos results represent the first measurement in the forward region. 

\section*{Acknowledgements}
%
%
\noindent We express our gratitude to our colleagues in the CERN
accelerator departments for the excellent performance of the LHC. We
thank the technical and administrative staff at the LHCb
institutes.
We acknowledge support from CERN and from the national agencies:
CAPES, CNPq, FAPERJ and FINEP (Brazil); 
MOST and NSFC (China); 
CNRS/IN2P3 (France); 
BMBF, DFG and MPG (Germany); 
INFN (Italy); 
NWO (Netherlands); 
MNiSW and NCN (Poland); 
MEN/IFA (Romania); 
MICINN (Spain); 
SNSF and SER (Switzerland); 
NASU (Ukraine); 
STFC (United Kingdom); 
DOE NP and NSF (USA).
We acknowledge the computing resources that are provided by CERN, IN2P3
(France), KIT and DESY (Germany), INFN (Italy), SURF (Netherlands),
PIC (Spain), GridPP (United Kingdom), 
CSCS (Switzerland), IFIN-HH (Romania), CBPF (Brazil),
Polish WLCG  (Poland) and NERSC (USA).
We are indebted to the communities behind the multiple open-source
software packages on which we depend.
Individual groups or members have received support from
ARC and ARDC (Australia);
Minciencias (Colombia);
AvH Foundation (Germany);
EPLANET, Marie Sk\l{}odowska-Curie Actions and ERC (European Union);
A*MIDEX, ANR, IPhU and Labex P2IO, and R\'{e}gion Auvergne-Rh\^{o}ne-Alpes (France);
Key Research Program of Frontier Sciences of CAS, CAS PIFI, CAS CCEPP, 
Fundamental Research Funds for the Central Universities, 
and Sci. \& Tech. Program of Guangzhou (China);
GVA, XuntaGal, GENCAT and Prog.~Atracci\'on Talento, CM (Spain);
SRC (Sweden);
the Leverhulme Trust, the Royal Society
 and UKRI (United Kingdom).

\clearpage
\newcommand{\xx}{\ensuremath{\kern 0.5em }}
\newcommand{\xxx}{\ensuremath{\kern 0.75em }}
\clearpage
\section*{Appendices}

\appendix

\section{Numerical results}
\label{sec:A}

\begin{table}[htbp]
    \caption{The differential cross-section for coherent \jpsi production as a function of $y^*$.}
  \label{tab:jpsi_cross}
      \begin{center}
      \begin{tabular}{cccccc}
      \hline
      Interval  &  $\deriv{\sigma}^\coh_{\jpsi}/\deriv y^*$ \unitmbarn & \multicolumn{4}{c}{Uncertainties \unitmbarn} \\
     \hline   
 & & Stat. & Syst. &  Lumi. & Total \\
             \cline{3-6}
$2.0<y^*<2.5$ & 3.392 & 0.108 & 0.165 & 0.147 & 0.247\\
$2.5<y^*<3.0$ & 2.896 & 0.053 & 0.117 & 0.127 & 0.181\\
$3.0<y^*<3.5$ & 2.395 & 0.040 & 0.089 & 0.105 & 0.144\\
$3.5<y^*<4.0$ & 1.922 & 0.039 & 0.072 & 0.084 & 0.117\\
$4.0<y^*<4.5$ & 1.181 & 0.054 & 0.049 & 0.052 & 0.089\\
\hline
$2.0<y^*<4.5$ & 5.965 & 0.059 & 0.232 & 0.262 & 0.355\\
\hline
    \end{tabular}
    \end{center}
  \end{table}

\begin{table}[htbp]
    \caption{The differential cross-section of coherent \psitwos production as a function of $y^*$.}
  \label{tab:psi2s_cross}
      \begin{center}
      \begin{tabular}{cccccc}
      \hline
      Interval  &  $\deriv{\sigma}^\coh_{\psitwos}/\deriv y^*$ \unitmbarn & \multicolumn{4}{c}{Uncertainties \unitmbarn} \\
     \hline   
 & & Stat. & Syst. &  Lumi. & Total \\
             \cline{3-6}
$2.0<y^*<2.5$ & 0.656 & 0.155 & 0.024 & 0.029 & 0.160\\
$2.5<y^*<3.0$ & 0.430 & 0.082 & 0.014 & 0.019 & 0.085\\
$3.0<y^*<3.5$ & 0.373 & 0.060 & 0.012 & 0.016 & 0.064\\
$3.5<y^*<4.0$ & 0.268 & 0.057 & 0.009 & 0.012 & 0.059\\
$4.0<y^*<4.5$ & 0.184 & 0.071 & 0.007 & 0.008 & 0.072\\
\hline
$2.0<y^*<4.5$ & 0.923 & 0.086 & 0.028 & 0.040 & 0.099\\
\hline
    \end{tabular}
    \end{center}
  \end{table}

\begin{table}[htbp]
\caption{The differential cross-section ratio between \psitwos and \jpsi coherent production as a function of $y^*$. The uncertainty due to luminosity determination cancels in the ratio.}
\label{tab:ratio_cross}
\begin{center}
\begin{tabular}{ccccc}
\hline
Interval  &   $\deriv{\sigma}^\coh_{\psitwos}/\deriv{\sigma}^\coh_{\jpsi}$ & \multicolumn{3}{c}{Uncertainties} \\
\hline   
 & & Stat. & Syst. & Total \\
 \cline{3-5}
$2.0<y^*<2.5$ & 0.193 & 0.046 & 0.004 & 0.046\\
$2.5<y^*<3.0$ & 0.149 & 0.028 & 0.003 & 0.029\\
$3.0<y^*<3.5$ & 0.156 & 0.025 & 0.003 & 0.026\\
$3.5<y^*<4.0$ & 0.139 & 0.030 & 0.003 & 0.030\\
$4.0<y^*<4.5$ & 0.156 & 0.061 & 0.003 & 0.061\\
\hline
$2.0<y^*<4.5$ & 0.155 & 0.014 & 0.003 & 0.015\\
\hline
    \end{tabular}
    \end{center}
  \end{table}

  \begin{table}[htbp]
  \def\pz{\phantom{0}}
    \caption{The double differential cross-section of coherent \jpsi production as a function of $\pt^*$ in rapidity range $2.0<y^*<4.5$.}
  \label{tab:jpsipt_cross_245}
      \begin{center}
      \begin{tabular}{rccccc}
      \hline
      Interval $[\mevc]$  &  $\deriv^2{\sigma}^\coh_{\jpsi}/\deriv \pt^*\deriv y^*$ \unitmbarngevc & \multicolumn{4}{c}{Uncertainties \unitmbarngevc} \\
     \hline   
 & & Stat. & Syst. &  Lumi. & Total \\
             \cline{3-6}
$0<\pt^*<\pz20$ & 13.391 & 0.352 & 0.908 & 0.587 & 1.138\\
$20<\pt^*<\pz40$ & 33.940 & 0.556 & 2.007 & 1.489 & 2.560\\
$40<\pt^*<\pz60$ & 35.077 & 0.495 & 1.462 & 1.538 & 2.179\\
$60<\pt^*<\pz80$ & 22.645 & 0.381 & 0.492 & 0.993 & 1.172\\
$80<\pt^*<100$ & 9.945 & 0.249 & 0.472 & 0.436 & 0.689\\
$100<\pt^*<120$ & \pz2.028 & 0.128 & 0.311 & 0.089 & 0.347\\
$120<\pt^*<140$ & \pz0.432 & 0.083 & 0.138 & 0.019 & 0.163\\
$140<\pt^*<160$ & \pz0.781 & 0.103 & 0.273 & 0.034 & 0.293\\
$160<\pt^*<180$ & \pz0.986 & 0.118 & 0.213 & 0.043 & 0.247\\
$180<\pt^*<200$ & \pz0.464 & 0.102 & 0.080 & 0.020 & 0.131\\
\hline
$0<\pt^*<200$ & 11.904 & 0.103 & 0.233 & 0.522 & 0.581\\
\hline
    \end{tabular}
    \end{center}
  \end{table}

 \begin{table}[]
   \def\pz{\phantom{0}}
    \caption{The double differential cross-section of coherent \psitwos production as a function of $\pt^*$ in rapidity range $2.0<y^*<4.5$.}
  \label{tab:psi2spt_cross_245}
      \begin{center}
      \begin{tabular}{rccccc}
      \hline
      Interval $[\mevc]$  &  $\deriv^2{\sigma}^\coh_{\psitwos}/\deriv \pt^* \deriv y^*$ \unitmbarngevc & \multicolumn{4}{c}{Uncertainties \unitmbarngevc} \\
     \hline   
 & & Stat. & Syst. &  Lumi. & Total \\
             \cline{3-6}
$0<\pt^*<\pz30$ & 2.073 & 0.942 & 0.141 & 0.091 & 0.957\\
$30<\pt^*<\pz70$ & 5.447 & 0.775 & 0.254 & 0.239 & 0.850\\
$70<\pt^*<\pz90$ & 3.476 & 0.535 & 0.110 & 0.152 & 0.567\\
$90<\pt^*<110$ & 1.136 & 0.337 & 0.108 & 0.050 & 0.357\\
$110<\pt^*<150$ & 0.000 & 0.093 & 0.000 & 0.000 & 0.093\\
$150<\pt^*<200$ & 0.025 & 0.051 & 0.006 & 0.001 & 0.051\\
\hline
$0<\pt^*<200$ & 1.833 & 0.160 & 0.052 & 0.080 & 0.187\\
\hline
    \end{tabular}
    \end{center}
  \end{table}
 
\clearpage
\addcontentsline{toc}{section}{References}
\bibliographystyle{LHCb}
\bibliography{main,standard,LHCb-PAPER,LHCb-CONF,LHCb-DP,LHCb-TDR}

\newpage
\centerline
{\large\bf LHCb collaboration}
\begin
{flushleft}
\small
R.~Aaij$^{32}$\lhcborcid{0000-0003-0533-1952},
A.S.W.~Abdelmotteleb$^{50}$\lhcborcid{0000-0001-7905-0542},
C.~Abellan~Beteta$^{44}$,
F.~Abudin{\'e}n$^{50}$\lhcborcid{0000-0002-6737-3528},
T.~Ackernley$^{54}$\lhcborcid{0000-0002-5951-3498},
B.~Adeva$^{40}$\lhcborcid{0000-0001-9756-3712},
M.~Adinolfi$^{48}$\lhcborcid{0000-0002-1326-1264},
H.~Afsharnia$^{9}$,
C.~Agapopoulou$^{13}$\lhcborcid{0000-0002-2368-0147},
C.A.~Aidala$^{76}$\lhcborcid{0000-0001-9540-4988}
% ...

{\footnotesize \it

% ...
$^{9}$Universit{\'e} Clermont Auvergne, CNRS/IN2P3, LPC, Clermont-Ferrand, France\\
% ...
$^{13}$LPNHE, Sorbonne Universit{\'e}, Paris Diderot Sorbonne Paris Cit{\'e}, CNRS/IN2P3, Paris, France\\
% ...
$^{32}$Nikhef National Institute for Subatomic Physics, Amsterdam, Netherlands\\
% ...
$^{40}$Instituto Galego de F{\'\i}sica de Altas Enerx{\'\i}as (IGFAE), Universidade de Santiago de Compostela, Santiago de Compostela, Spain\\
% ...
$^{44}$Physik-Institut, Universit{\"a}t Z{\"u}rich, Z{\"u}rich, Switzerland\\
% ...
$^{48}$H.H. Wills Physics Laboratory, University of Bristol, Bristol, United Kingdom\\
% ...
$^{50}$Department of Physics, University of Warwick, Coventry, United Kingdom\\
% ...
$^{54}$Oliver Lodge Laboratory, University of Liverpool, Liverpool, United Kingdom\\
% ...
$^{62}$Syracuse University, Syracuse, NY, United States\\
% ...
$^{76}$University of Michigan, Ann Arbor, MI, United States, associated to $^{62}$\\
\bigskip
% ...
}
\end{flushleft}

\end{document}